\documentclass{article}
\begin{document}

\title{Flat histogram Monte Carlo method}

\author{Jian-Sheng Wang,\\
Department of Computational Science,\\
National University of Singapore,\\
Singapore 119260,\\
Republic of Singapore}

\date{9 September 1999}
\maketitle

\begin{abstract}

We discuss a sampling algorithm which generates flat
histogram in energy.  In combination with transition matrix
Monte Carlo, the density of states and derived quantities such
as entropy and free energy as a function of temperature can be
computed in a single simulation.

\end{abstract}

\section{Introduction}

The basic problem in equilibrium statistical mechanics is to
compute the canonical average 
\begin{equation}
  \langle A \rangle_T = { \sum_{\{\sigma\}} A(\sigma) 
        \exp\bigl(- H(\sigma)/kT\bigr)
  \over \sum_{\{\sigma\}} \exp\bigl(- H(\sigma)/kT\bigr) } .
\end{equation}
In addition, free energy 
\begin{equation}
F = - kT \ln \sum_{\{\sigma\}} \exp\bigl( - H(\sigma)/kT\bigr)
\end{equation}
and related quantity entropy are also very important.  Standard
Monte Carlo method, e.g., Metropolis importance sampling
algorithm, is simple and general.  However, computation of free
energy \cite{free-energy-review} is difficult with such method. 

Over the last decade, there have been a number of methods
addressing this problem
\cite{FS-histo,Berg,Lee,Oliveira,Wang-TMMC}.  A common theme in
these approaches is to evaluate the density of states $n(E)$
directly.  If this can be done with sufficient accuracy, then
the summation over all configuration states can be rewritten as
sum over energy only, e.g.,
\begin{equation}
F = - kT \ln \sum_{E} n(E) \exp(-E/kT).
\end{equation}
Can we evaluate $n(E)$ with a uniform relative accuracy for all
$E$?  Our experience with flat histogram method suggests a
positive yes.

\section{Broad histogram equation}

Oliveira et al \cite{Oliveira} showed the validity of following
equation relating density of states with the microcanonical
average number of moves in a single spin flip dynamics:
\begin{equation}
  n(E) \langle N(\sigma, \Delta E) \rangle_E = 
  n(E+ \Delta E) \langle N(\sigma', - \Delta E) \rangle_{E+\Delta E}.
\label{eq-bhe}
\end{equation}
This equation is equivalent to a detail balance condition in an
infinite temperature transition matrix Monte Carlo simulation
\cite{Wang-TMMC}.  The quantity $N(\sigma, \Delta E)$ is the
number of ways that the system goes to a state with energy $E +
\Delta E$ by a single spin flip, given that the current state
is $\sigma$ with energy $E$.  The average $\langle \cdots
\rangle_E$ is performed over all the states with a fixed
initial energy $E$ (i.e., a microcanonical average).

\section{A flat histogram dynamics} 
Consider the follow Monte Carlo dynamics \cite{flat-histo}
\begin{enumerate}
 \item Pick a site at random.
 \item Flip the spin with probability $r(E'|E)$.
 \item Sample $N(\sigma, \Delta E)$, i.e., accumulate the statistics
       for $\langle N(\sigma, \Delta E) \rangle$.
 \item Go to 1.
\end{enumerate} 
The flip probability $r$ is given by
\begin{equation}
   r(E'|E) = \min\left(1, { \langle N(\sigma', -\Delta E) \rangle_{E'} 
                \over \langle N(\sigma, \Delta E) \rangle_E } \right),
\end{equation} 
where the current state $\sigma$ has energy $E$, and new state
$\sigma'$ with one spin flipped has energy $E'=E + \Delta E$.

With the above choice of flip rate, we can show that detail
balance is satisfied
\begin{equation}
   r(E'|E) P(\sigma) = r(E|E') P(\sigma')
\end{equation}
if $P(\sigma) = {\rm const} /n(E(\sigma))$, since this equation
is equivalent to the broad histogram equation,
Eq.~(\ref{eq-bhe}).  The histogram in energy is then
$H(E) \propto \sum P(\sigma) = {\rm constant}$.
It turns out that the choice of flip rate is not unique; many
other formulas are possible \cite{Swendsen-Wang-Li-et-al}.

Due to Eq.~(\ref{eq-bhe}), the flip rate is also equal to
$\min[1, n(E)/n(E')]$.  This is exactly Lee's method \cite{Lee}
of entropy sampling (which is equivalent to multicanonical method
\cite{Berg}).  However, since neither $n(E)$ nor $\langle
N(\sigma, \Delta E) \rangle_E$ is known before simulation, the
way by which simulation gets boot-trapped is quite different.  Our
method is very efficient in this respect.  Another important
difference is that we take $\langle N(\sigma, \Delta E)
\rangle_E$ as our primary statistics, from which we derive the
density of states $n(E)$.  Apart from the number of iterations
needed, the transition matrix results are in general more
accurate than that obtained using the energy histogram
\cite{Lima}.

\section{Simulation procedures}

The first approximate scheme we use is to replace the true
microcanonical average by a cumulative sample average
\begin{equation}
   \langle N(\sigma, \Delta E) \rangle_E \approx 
   { 1 \over M } \sum_{i=1}^M N(\sigma^i, \Delta E),
\end{equation}
where the samples $\sigma^i$ are configurations 
generated during simulation with
energy $E$.  Each state in the
simulation contributes to $N(\cdots)$ for some $E$.  
For those $E'$ where data are not available, we set
$r=1$.  This biases towards unvisited energy states. This
dynamics does not satisfy detail balance exactly since the
transition rate is fluctuating.  However, test on small systems
shows that the sample averages do converge to the exact value
with errors proportional to the inverse of square root of Monte
Carlo steps.

A two-stage simulation will guarantee detail balance.  Stage
one is the same as described above.  In stage two, we adjust
the approximate transition matrix obtained in stage one such
that detail balance is satisfied exactly.  In this stage, the
flip rate is fixed and will not fluctuate with the simulation.
The second stage simulation is dynamically equivalent to Berg's
multicanonical or Lee's entropy sampling dynamics.  The stage
two can be iterated so that the simulated ensemble approaches
the ideal multicanonical ensemble,  but we found two-stage or
even a single stage simulation already gives excellent results.

The simulation can also be combined with the N-fold way
\cite{N-fold} with little overhead in computer time, since the
quantity $N(\sigma, \Delta E)$ needed in the N-fold way is
already computed.  In addition, not only we can have equal
histogram (multicanonical), we can also generate ``equal-hit''
ensemble, where each energy of distinct states is visited
equally likely
\cite{Swendsen-Wang-Li-et-al}. 

\section{Density of states from transition matrix}

The density of states is related to the transition matrix
$T_{E,\Delta E} = \langle N(\sigma, \Delta E) \rangle_E/N$
by Eq.~(\ref{eq-bhe}),
where $N$ is the total number of possible moves.  Since there
are more equations than unknown $n(E)$, we use a least-square
method to obtain ``optimal'' solution.  Let $S(E) = \ln n(E)$,
we consider
\begin{equation}
   {\rm minimize} \quad 
   \sum_{E, E'}  {1 \over \sigma^2_{E, E'}} \left( 
    S(E')  - S(E) - \ln { T_{E, \Delta E} \over T_{E', -\Delta E} } 
   \right)^2
\end{equation} 
subject to all the conditions known.  For example, for the Ising
model, we have $n(E_{min}) = n(E_{max}) = 2$, and $\sum_E n(E)
= 2^N$.  The variance $\sigma^2$ is the variance of the
quantity $\ln { T_{E, \Delta E} / T_{E', -\Delta E} }$ obtained
from sets of Monte Carlo data.  It is also possible to work
with the matrix $T$ directly with the conditions that $T$ must
satisfy.

\section{Conclusion} 

We proposed an algorithm which samples energy $E$ uniformly.
Comparing to multicanonical simulation, the method offers a
very easy way of starting the simulation.  The dynamic
characteristics are similar to well-converged multicanonical
Monte Carlo dynamics.  For example, the tunneling time for
10-state Potts model in two dimensions is about $\tau \propto
L^{2.6}$ for $L \times L$ system.  
It is very easy to combine statistics from several
simulations, including parallel simulations.  It is an
efficient method for computing density of states and all the
related thermodynamic quantities.

\section*{Acknowledgements}

The work presented here is in collaborations with 
R.H. Swendsen, T.K. Tay, L.W. Lee, and Z.F. Zhen.

\end{document}